\documentclass[conference]{IEEEtran}

\usepackage[T1]{fontenc}
\usepackage{graphicx}
\usepackage[hidelinks]{hyperref}
\usepackage{float}
\usepackage{algorithm}
\usepackage{algpseudocode}
\usepackage{tabularx}
\usepackage{booktabs}
\usepackage{subcaption}
\usepackage{amsmath}
\usepackage{multirow}
\usepackage{longtable}
\usepackage{listings}
\usepackage[dvipsnames]{xcolor}

\makeatletter
\def\section{\@startsection{section}{1}{\z@}%
  {1.5ex plus 1.5ex minus .5ex}%
  {0.7ex plus 1ex minus 0ex}%
  {\normalfont\normalsize\bfseries\centering}}
\def\subsection{\@startsection{subsection}{2}{\z@}%
  {1.5ex plus 1.5ex minus .5ex}%
  {0.7ex plus .5ex minus 0ex}%
  {\normalfont\normalsize\bfseries\itshape}}
\makeatother

\begin{document}
\raggedbottom

\title{Autotuning GPU Thread Block Sizes for Large-Scale Stencil Applications Using the OPS DSL}

\author{
\IEEEauthorblockN{Kevin Antonio Nava Garcia\IEEEauthorrefmark{1},
Istvan Zoltan Reguly\IEEEauthorrefmark{1}}
\IEEEauthorblockA{\IEEEauthorrefmark{1}
Faculty of Information Technology and Bionics, Pázmány Péter Catholic University, Budapest, Hungary\\
Email: \{nava.garcia.kevin.antonio, reguly.istvan\}@itk.ppke.hu}
}

\maketitle

 \begin{abstract}
  Large-scale structured-mesh stencil applications can run for hours or days on
  modern HPC GPU platforms, making launch-configuration choices important. OPS
  enables high-level stencil codes to target multiple GPU architectures, but its
  GPU backends rely on a default or user-selected global block/workgroup shape,
  even though different kernels, problem geometries, and architectures may favor
  different configurations.

  This paper presents a transparent per-kernel block/workgroup-shape autotuning
  extension for OPS. The extension explores valid block/workgroup shapes during
  execution, records the fastest configuration for each kernel, and enables
  precomputed replay of the selected configurations in subsequent runs.

  We evaluate the approach on structured-mesh proxy and academic research
  applications, covering 459 unique kernels across 3 NVIDIA and 2 AMD GPUs.
  The evaluation reports speedup distributions, representative wall-time
  summaries, timing variability, and input problem shape effects. Replaying the
  best per-kernel launch shapes improves 923 of 995 measured
  application/input/GPU cases over the default OPS shape. Depending on the GPU,
  the median speedup ranges from 1.02$\times$ to
  1.18$\times$, with individual configurations of LAPL2D, TTI3D,
  and OpenSBLI-based TGV solvers exceeding 1.5$\times$. The results also show
  that performance sensitivity is not limited to launch shape. At fixed cell
  count, changing the input problem shape changes precomputed-replay runtime by
  0.8--34.6\% in representative cases, while kernel-level timing variability
  depends on the interaction between runtime contribution, input problem shape,
  and GPU/backend behavior. Overall, automatic per-kernel
  block/workgroup-shape selection improves OPS performance across GPUs, provided
  shape sensitivity and timing stability are interpreted together.

\end{abstract}

\section{Introduction}
  Structured-mesh stencil algorithms are widely used in scientific and engineering simulations, including computational fluid dynamics (CFD), heat transfer, shallow-water
  models, wave propagation, and image processing. These applications typically apply stencil-based computations over structured domains for many time steps. In this context, a kernel is the computation applied at each mesh point, for example
  one stencil update over all grid points. Since such kernels are executed many times in long-running simulations, even modest improvements to individual kernel executions can translate into substantial reductions in total
  application runtime.

  The regular structure of these applications makes them well suited to domain-specific programming frameworks. OPS (Oxford Parallel Library for Structured Mesh Solvers) is one such framework: it is designed specifically for applications operating on multi-block structured grids
  and using explicit solvers~\cite{ops-paper}. Rather than requiring developers to write backend-specific GPU or CPU code, OPS lets them describe stencil kernels at a high level. The OPS translator and runtime then generate
  backend-specific implementations for CPUs, GPUs, and distributed-memory systems. This separation improves programmer productivity and performance portability: developers focus on the numerical method, while
  OPS manages backend code generation, data movement, halo exchanges, and kernel launches.

However, performance portability does not imply performance optimality. Even when the same OPS application can run across different
  GPU architectures, the performance of the generated code still depends on low-level execution parameters. Prior work has shown that
  GPU heterogeneity makes a single tuning choice unlikely to be optimal across devices and kernels, and identifies workgroup
  configuration as one possible direction for performance improvement~\cite{petrovic2020ktt}. In the OPS context, this directly
  motivates block/workgroup-shape tuning: different kernels and GPU backends may favor different launch configurations. OPS currently
  uses a default global block/workgroup shape applied uniformly across kernels, which can limit performance. This creates an
  opportunity for automatic, per-kernel block/workgroup-shape selection inside the OPS runtime.

 \subsection{Why This Remains Unsolved}

  Choosing an effective block/workgroup shape is deceptively simple. Many shapes
  are logically valid, but GPU threads execute in fixed-size groups, such as
  NVIDIA warps of 32 threads and AMD wavefronts of 64 threads~\cite{gpu_vendor_guides}.
  Poorly chosen shapes can leave lanes idle or restrict resident parallel work.
  They also affect occupancy: larger blocks may reduce the number of concurrent
  blocks per unit, while blocks that are too small may expose too little work.
  Maximum occupancy, however, does not guarantee best performance~\cite{Volkov2010Better}.

 The best block/workgroup shape is not universal. It depends
  on the kernel, the iteration-space geometry, the stencil access
  pattern, and the target GPU architecture. A shape that performs
  well for one kernel may be suboptimal for another kernel in the
  same application, or for the same kernel on a different GPU.
  This is particularly relevant for structured-mesh applications,
  where many different kernels may operate over different
  ranges, dimensions, and data access patterns. Zhang and
  Mueller~\cite{zhang2013stencilautotuning} made a related
  observation for generated 3D stencil codes, showing that the
  best block-related configuration depends on the stencil pattern
  and GPU architecture. While their tuning target is the generated
  stencil code, our tuning target is the OPS-generated kernel.

  Manual tuning does not scale to this setting. A single global block/workgroup
  shape cannot capture per-kernel differences, yet repeated profiling distracts
  application developers from numerical correctness and model development. OPS is
  a natural place to automate this choice because it separates user-written
  stencil kernels from backend-specific generated code. We therefore integrate
  per-kernel block/workgroup-shape selection inside the OPS runtime and generated
  GPU wrappers, enabling tuning without changes to application code.

\subsection{Contributions}

In this paper, we make the following contributions:
\begin{itemize}

\item A transparent per-kernel block/workgroup-shape autotuning mechanism integrated into the OPS runtime and generated GPU backend wrappers.

\item A replay mode that applies selected per-kernel shapes from a prior tuning run, isolating their performance effect from online search overhead.

\item A comprehensive evaluation across 8 structured-mesh applications, 459 unique kernels, and 3 NVIDIA and 2 AMD GPUs, including speedup distributions, representative wall times, timing variability, and input problem shape sensitivity.

\end{itemize}

\section{Related Work}

Autotuning is a well-established strategy for recovering
  performance portability by empirically selecting implementation
  or execution parameters for a platform. ATLAS~\cite{whaley2001atlas}
  demonstrated this idea for dense linear algebra routines in
  BLAS, while OSKI~\cite{vuduc2005oski} applied automatic tuning
  to sparse matrix kernels such as sparse matrix-vector multiply
  and sparse triangular solve. ATLAS selects among generated or
  alternative BLAS implementations and blocking/register/cache
  parameters; OSKI selects matrix-dependent data structures and
  transformations for sparse kernels. These works establish
  empirical tuning as a practical mechanism for adapting numerical
  software to both architecture and computation structure.

  In the GPU context, Tillmann et al.~\cite{tillmann2014} proposed AtuneRT, an
  online autotuner using Nelder--Mead to search CUDA launch parameters. Their
  method tunes scalar block size, expressed as total threads per block, and loop
  unrolling, reporting up to 33\% speedup over hand-picked defaults. However,
  scalar block-size tuning does not capture per-dimension block/workgroup shape
  $(b_x,b_y,b_z)$, which matters for structured meshes because equal thread
  counts can interact differently with geometry, stencil access, and memory layout.

\subsection{Kokkos Tuning Approaches}
Kokkos is a C++ performance-portability library that lets a single source program target multiple HPC backends (CUDA, HIP, SYCL, OpenMP) by abstracting parallel execution and memory access~\cite{kokkos2014}. Its autotuning subsystem exposes a callback protocol that delegates performance-sensitive parameters, such as tile dimensions, block/workgroup sizes, and algorithmic variants, to an external plugin loaded at runtime. Implementations such as APEX~\cite{huck2015apex}, built on Active Harmony~\cite{tiwari2009activeharmony}, and Tasmanian illustrate the range of search strategies this interface supports~\cite{kokkostuningdesign}. Kokkos requires users to adopt its programming model and integrate tuning callbacks; it does not generate kernel code on the user's behalf, so launch-parameter optimization remains an application-level concern. In contrast, OPS separates high-level user code from generated low-level implementation details, making automatic block/workgroup-shape selection possible below the user's level of abstraction.

\subsection{Kernel Tuning Toolkit KTT}

Kernel Tuning Toolkit (KTT)~\cite{petrovic2020ktt} autotunes CUDA and OpenCL kernels by exploring programmer-defined parameters, such as work-group size, tiling, memory placement, thread
  coarsening, and loop unrolling. These parameters must be exposed through KTT, which manages kernel compilation, execution, and measurement. In contrast, our work transparently tunes block/
  workgroup shapes within the OPS-generated GPU backend, without requiring user-written kernels to be parametrized, rewritten, or annotated. Thus, KTT provides application-level kernel tuning,
  whereas our approach hides launch-shape tuning beneath the OPS application interface.

\subsection{Auto-Generation and Auto-Tuning of 3D Stencil Codes on
GPU Clusters}

 Zhang and Mueller~\cite{zhang2013stencilautotuning} present a framework that generates and autotunes 3D stencil codes for GPU clusters from compact stencil specifications. Their exhaustive search
  considers logical block sizes, CUDA thread-block dimensions, and texture-memory usage to select the best valid configuration for each stencil and GPU. While both approaches exploit the sensitivity
  of stencil computations to GPU launch configurations, their work targets generated CUDA stencil implementations, whereas ours tunes OPS-generated kernels across complete applications and multiple
  GPU backends.

\subsection{Kernel Tuner}

 Kernel Tuner~\cite{werkhoven2019kerneltuner} is a Python framework for
  autotuning CUDA, OpenCL, and C kernels over user-exposed parameters such as
  block dimensions, tiling, and unrolling, with search strategies including
  brute force, random sampling, basin hopping, and differential evolution. In
  contrast, our work requires no external scripts or manual parameterization of
  generated kernels; tuning is integrated into the OPS runtime/backend layer and
  records selected per-kernel shapes for replay.

\subsection{IterML : Iterative Machine Learning for Intelligent Parameter Pruning and Tuning in Graphics Processing Units}

 Cui and Feng~\cite{cui2021iterml} proposed IterML, an iterative machine-learning framework for pruning the parameter search space and tuning GPU thread-block dimensions. By predicting configuration performance, IterML reduces the effort required to find near-optimal settings. This approach aligns with our future work on intelligent parameter tuning, although the framework and benchmarks differ of our future work.

\section{Background}

This section introduces the OPS structure, code reference, and integration point
for the tuning algorithm.

\subsection{OPS and Autotuning framework}

\begin{figure}[H]
        \centering
        \includegraphics[width=0.8\columnwidth]{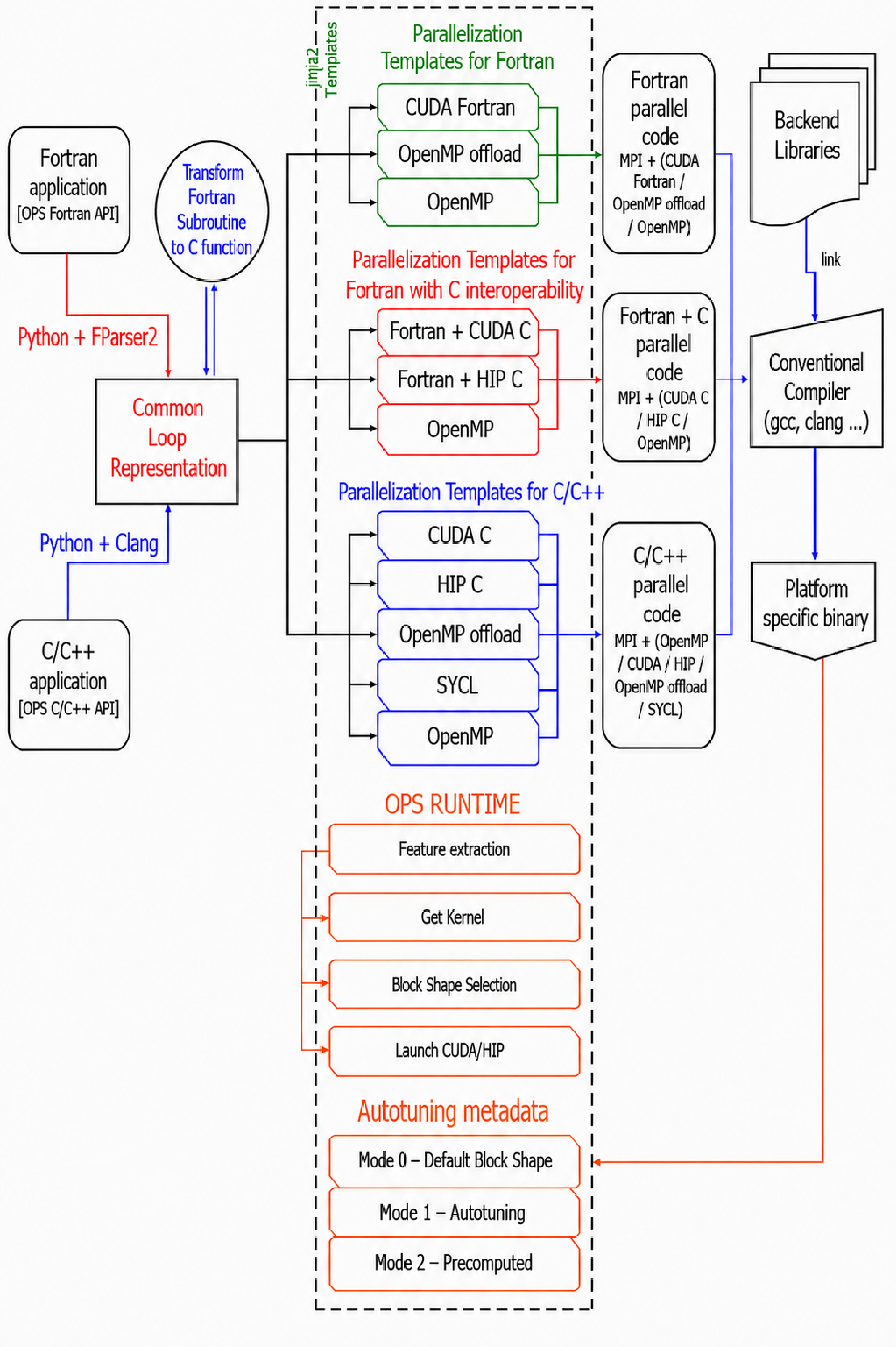}
        \caption{OPS source-to-backend workflow with runtime block/workgroup-shape tuning.}
        \label{fig:ops-workflow-tuning}
  \end{figure}

Figure~\ref{fig:ops-workflow-tuning} summarizes the overall OPS workflow used in
  this work, from the high-level application description to backend generation,
  runtime block/workgroup-shape selection, and autotuning metadata management.

 \subsection{OPS Kernel and Launch Metadata}

  \definecolor{opsCodeKeyword}{RGB}{0,0,180}
  \definecolor{opsCodeComment}{RGB}{0,140,0}
  \definecolor{opsCodeString}{RGB}{160,70,0}
  \definecolor{opsCodeRule}{RGB}{145,145,145}
  \definecolor{opsCodeBg}{RGB}{250,250,250}

  \lstdefinestyle{opside}{
    language=C++,
    basicstyle=\scriptsize\ttfamily,
    keywordstyle=\color{opsCodeKeyword}\bfseries,
    commentstyle=\color{opsCodeComment},
    stringstyle=\color{opsCodeString},
    backgroundcolor=\color{opsCodeBg},
    frame=tb,
    rulecolor=\color{opsCodeRule},
    framerule=0.4pt,
    xleftmargin=0.4em,
    xrightmargin=0.4em,
    aboveskip=0.25em,
    belowskip=0.25em,
    columns=fullflexible,
    keepspaces=true,
    showstringspaces=false,
    breaklines=true,
    tabsize=2,
    morekeywords={
      ops_init,ops_exit,ops_partition,ops_decl_block,ops_decl_dat,
      ops_decl_stencil,ops_par_loop,ops_arg_dat,ops_arg_gbl,
      ops_block,ops_dat,ops_stencil,OPS_READ,OPS_WRITE,
      OPS_ACC0,OPS_ACC1,dim3
    }
  }

  An OPS application is built around an explicit data model for structured mesh
  computations, rather than backend-specific parallel code. The application
  declares structured blocks, datasets, and stencil access patterns; OPS uses this
  metadata to manage memory layout, memory spaces, halo regions, data movement,
  and MPI domain decomposition~\cite{ops-paper}. Parallel execution is expressed
  through \texttt{ops\_par\_loop}, which applies a user-written single-point
  kernel over an iteration range and binds it to datasets, stencils, and access
  modes.

  Listing~\ref{lst:ops-user-kernel} shows a simplified OPS kernel and its
  corresponding parallel-loop invocation. The function \texttt{apply\_stencil}
  defines the point-wise computation, while the \texttt{ops\_par\_loop} call
  describes how that computation is applied over the structured grid.

  \begin{lstlisting}[style=opside,
  caption={Simplified OPS user kernel and parallel-loop invocation.},
  label={lst:ops-user-kernel}]
  void apply_stencil(const ACC<double>& A, ACC<double>& Anew)
  {
    Anew(0,0) = 0.25 * (
        A( 1, 0) + A(-1, 0) +
        A( 0, 1) + A( 0,-1));
  }

  int range[] = {1, nx - 1, 1, ny - 1};

  ops_par_loop(apply_stencil, "apply_stencil", block, 2, range,
    ops_arg_dat(A,    1, S2D_5PT, "double", OPS_READ),
    ops_arg_dat(Anew, 1, S2D_00,  "double", OPS_WRITE));
  \end{lstlisting}

  This call exposes the metadata used by OPS to generate backend code. Here,
  \texttt{A} is read through \texttt{S2D\_5PT}, \texttt{Anew} is written through
  \texttt{S2D\_00}, and the range defines the grid region. More generally, each
  \texttt{ops\_par\_loop} identifies the kernel, dimensionality, iteration space,
  datasets, stencil descriptors, and access modes.

  The source-to-source translator then generates OpenMP, CUDA, HIP, SYCL, and
  MPI-enabled variants. For GPU backends, the generated host wrapper computes
  the local range, selects a block/workgroup shape, constructs the launch grid,
  and invokes the device kernel. Listing~\ref{lst:ops-generated-launch} shows
  the launch point relevant to this work.

  \begin{lstlisting}[style=opside,
  caption={Generated GPU launch path where the block/workgroup shape is selected before launching the generated kernel.},
  label={lst:ops-generated-launch}]
  dim3 block = ops_get_kernel_block_size(
    kernel_id, dim, local_range, nargs, args,
    maxThreadsPerBlock, numRegs);

  dim3 grid((range_x - 1) / block.x + 1,
            (range_y - 1) / block.y + 1);

  ops_apply_stencil<<<grid, block>>>(...);
  \end{lstlisting}

  This generated launch path is the baseline mechanism considered here. The
  autotuning extension changes only how the block/workgroup shape is selected at
  this point; it does not modify the OPS data model, user-written kernels, or
  \texttt{ops\_par\_loop} interface.

\subsection{Applications}

We evaluate 2D and 3D OPS applications from structured-mesh domains, including
  CloverLeaf, a compressible hydrodynamics proxy application ported to
  OPS~\cite{mudalige2014cloverleaf}, and SENGA2, a high-order finite-difference
  turbulent reacting-flow solver re-engineered with OPS~\cite{cant2012senga2}.
  The suite also covers OpenSBLI-based Taylor-Green vortex
  solvers~\cite{lusher2021opensbli}, TTI3D for seismic wave propagation, LBM2D,
  a D2Q9 lattice Boltzmann solver, and LAPL2D, a two-dimensional
  Laplace/Poisson stencil solver. These applications span 2D and 3D domains,
  memory-access patterns, computational intensities, stencil patterns, and
  kernel counts from a few kernels to more than one hundred generated parallel
  loops per application.

  This diversity motivates using individual generated kernels, rather than the
  full application, as the unit for block/workgroup-shape selection.

   \begin{table}[H]
    \centering
    \scriptsize
    \caption{Applications used in the evaluation. All applications are implemented in OPS and execute structured-mesh stencil kernels.}
    \label{tab:app_kernels}
    \setlength{\tabcolsep}{2.5pt}
    \begin{tabular}{@{}lllr@{}}
    \toprule
    \textbf{App} & \textbf{Dim.} & \textbf{Problem / method} & \textbf{Kernels} \\
    \midrule
    CL3D   & 3D & Explicit compressible hydrodynamics & 140 \\
    SENGA2 & 3D & Turbulent reacting-flow combustion CFD & 102 \\
    TGV-SA & 3D & Compressible Taylor-Green vortex, derivatives stored & 86 \\
    TGV-SN & 3D & Compressible Taylor-Green vortex, derivatives recomputed & 26 \\
    TTI3D  & 3D & Finite-difference seismic wave propagation & 3 \\
    CL2D   & 2D & Explicit compressible hydrodynamics & 82 \\
    LBM2D  & 2D & Explicit lattice Boltzmann D2Q9 method & 15 \\
    LAPL2D & 2D & Iterative Laplace/Poisson stencil solver & 5 \\
    \bottomrule
    \end{tabular}
  \end{table}

\section{Baseline and Autotuning Design in OPS}
\label{sec:autotuning}

Before each generated GPU kernel launch, OPS calls a selector that maps the
kernel metadata and backend limits to a launch shape
$B=(b_x,b_y,b_z)$, with unused dimensions set to one. The selector dispatches to
the active mode and returns only the launch shape; the generated wrapper then
computes the launch grid as $G=\lceil R/B\rceil$ and launches the kernel once.

\begin{algorithm}[H]
  \caption{Global OPS Block/Workgroup-Shape Selection}
  \footnotesize
  \begin{algorithmic}[1]
    \Require kernel id $k$, dimensionality $d$, range $R$, arguments $A$,
    backend limits $L$
    \Ensure Launch shape $B$
    \State Read the autotuning mode $m$
    \If{$m = 0$}
      \State $B \gets \Call{DefaultBlock}{k,d,R,A,L}$
    \ElsIf{$m = 1$}
      \State $B \gets \Call{AutotuneBlock}{k,d,R,A,L}$
    \ElsIf{$m = 2$}
      \State $B \gets \Call{PrecomputedBlock}{k,d,R,A,L}$
    \EndIf
    \State Compute the GPU launch grid as $G \gets \lceil R/B\rceil$
    \State Launch the generated kernel using grid $G$ and launch shape $B$
    \State \Return $B$
  \end{algorithmic}
\end{algorithm}

\subsection{Mode 0: Default Block/Workgroup Shape}
\label{sec:manual_tuning}
The baseline mode uses the standard OPS launch path shown in Listing~\ref{lst:ops-generated-launch}. The default block/workgroup shape is $(32,4,1)$, corresponding to 128 GPU threads or work-items per launch block.

\subsection{Mode 1: Dynamic Autotuning Mode}
\label{sec:runtime_autotune}
Mode~1 maintains a per-kernel tuning state and tests one valid candidate shape
per kernel invocation until all candidates have been evaluated.

   \begin{algorithm}[H]
    \caption{Mode 1: Runtime Autotuning}
    \scriptsize
    \begin{algorithmic}[1]
      \Require kernel id $k$, dimensionality $d$, range $R$, backend limits $L$
      \Ensure Launch shape $B$
      \State Get tuning state $S$ and open the run-specific CSV log
      \If{a pending candidate from the previous invocation has elapsed time $t$}
        \State Append $(k,S.pending,t)$ to the CSV log
        \If{$t < S.best\_time$}
          \State Update $S.best \gets S.pending$ and $S.best\_time \gets t$
        \EndIf
        \If{all candidates tested} \State Mark $S$ decided and save $S.best$ \EndIf
      \EndIf
      \If{$S$ is decided} \State \Return $S.best$ \EndIf
      \State Build valid power-of-two set $C$ with $32 \leq b_xb_yb_z \leq \min(1024,L)$
      \State Add the default shape to $C$ if missing
      \State Select the next untested candidate $B \in C$ for this invocation
      \State Store $B$ as $S.pending$ for timing after the wrapper launch
      \State \Return $B$
    \end{algorithmic}
  \end{algorithm}
\subsection{Mode 2: Precomputed Replay}
\label{sec:precompute}
Mode~2 replays the best launch shapes recorded by Mode~1 before each generated
kernel launch. Entries are keyed by kernel identity, backend/GPU,
dimensionality, and normalized local range/input shape, distinguishing the same
kernel across geometries or devices. If no match is found, the runtime falls
back to the default shape; the same selector is used by the CUDA and HIP
wrappers.

\section{Evaluation methodology}

\begin{figure}[H]
    \centering
    \includegraphics[width=0.8\columnwidth]{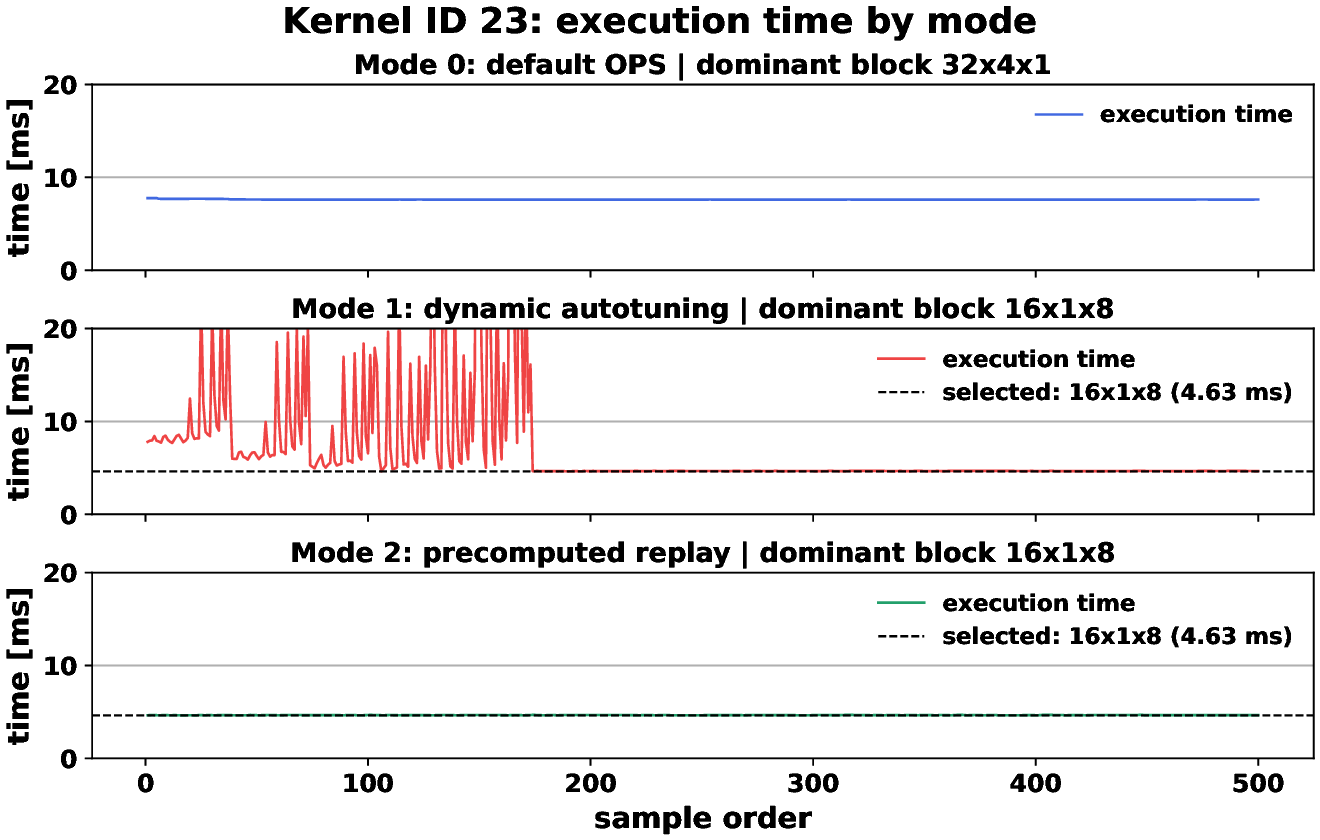}
     \caption{Kernel ID~23 timing on A100 across the three OPS selection modes.}
    \label{fig:mode-comparison-tgvstorenone}
  \end{figure}

 OPS exposes default execution, dynamic autotuning, and precomputed replay as
  separate measurement paths, separating online-search overhead from the
  performance of the selected block/workgroup shapes. Figure~\ref{fig:mode-comparison-tgvstorenone}
  illustrates this for kernel ID~23 on an NVIDIA A100: dynamic autotuning shows
  candidate-shape spikes before reusing the selected shape, while precomputed
  replay skips the search and applies that shape directly. For production-style
  runs with many solver iterations, the one-time Mode~1 exploration cost is
  amortized by reusing the selected shapes, so precomputed replay is the relevant
  performance path.

   We report speedup for each application and input configuration as
  \[
  S = \frac{T_{\mathrm{default}}}{T_{\mathrm{precomputed}}},
  \]
  where $T_{\mathrm{default}}$ uses the default OPS block/workgroup shape and
  $T_{\mathrm{precomputed}}$ replays the best shape selected for each kernel.

  Each configuration fixes the computational domain and solver iteration count.
  We execute each configuration five times to compute variability statistics and
  quantify measurement noise under the same application, input, GPU, and
  block/workgroup-selection mode.

\section{Evaluation and Setup}

Table~\ref{tab:app_kernels} summarizes the benchmark applications, while
Table~\ref{tab:problem-configs} reports the input configurations and iteration
counts used for repeated measurements. All benchmarks are OPS implementations.

\subsection{Experimental Setup and Benchmark}

The experiments were performed on five GPU platforms: NVIDIA V100 16 GB,
NVIDIA A100 40 GB, NVIDIA H100 80 GB on an x86 host system, AMD Instinct
MI300A 128 GB nominal HBM, and AMD Instinct MI250 128 GB using one GCD. The MI250
experiments were run on LUMI, while the remaining platforms were evaluated on
separate systems.

Most applications use 30 input configurations: three problem sizes of
approximately 8, 27, and 64 million cells, with ten aspect-ratio variants per
size. SENGA2 uses five cubic 3D configurations on H100, MI300A, and A100 only;
the SENGA2 cases exceeded V100 memory, and MI250 SENGA2 runs are not included.
LBM2D has 20
configurations in the aggregate summaries, covering only its 8M- and 27M-cell
input groups. Each completed configuration was executed five times per GPU.
Iteration counts were reduced for benchmarking practicality, although
production runs can reach millions of iterations. Table~\ref{tab:problem-configs}
summarizes the input set.

  \begin{table}[H]
  \centering
  \footnotesize
  \caption{Problem configurations used in the experiments. Each completed configuration was repeated five times on each evaluated GPU platform.}
  \label{tab:problem-configs}
  \setlength{\tabcolsep}{3pt}
  \begin{tabular}{@{}llcc@{}}
  \toprule
  \textbf{Application} & \textbf{Grid type} & \textbf{Configs} & \textbf{Iterations/run} \\
  \midrule
  CL3D     & 3D, $\sim$8M, 27M, 64M cells   & 30 & 220 \\
  TGV-SA   & 3D, $\sim$8M, 27M, 64M cells   & 30 & 300 \\
  TGV-SN   & 3D, $\sim$8M, 27M, 64M cells   & 30 & 500 \\
  TTI3D    & 3D, $\sim$8M, 27M, 64M cells   & 30 & 1001 \\
  SENGA2   & 3D, $\sim$17M cells       & 5  & 100 \\
  CL2D     & 2D, $\sim$8M, 27M, 64M cells   & 30 & 300 \\
  LBM2D    & 2D, $\sim$8M and 27M cells   & 20 & 400 \\
  LAPL2D   & 2D, $\sim$8M, 27M, 64M cells   & 30 & 400 \\
  \bottomrule
  \end{tabular}
  \end{table}

\section{Results}

Overall, the experiments cover default OPS execution, dynamic autotuning with
online search, and precomputed replay across five GPU platforms. The aggregate
speedup and variability analyses compare default OPS execution against
precomputed replay; for each application, grid configuration, GPU, and analyzed
mode, we record five wall-time repeats, yielding 9,950 executions in that
dataset. The following sections summarize speedup distributions, representative
wall-time comparisons, timing variability, and input problem shape effects.

\subsection{Application-Level Speedup Overview}

\begin{figure}[H]
      \centering
      \includegraphics[width=1\columnwidth]{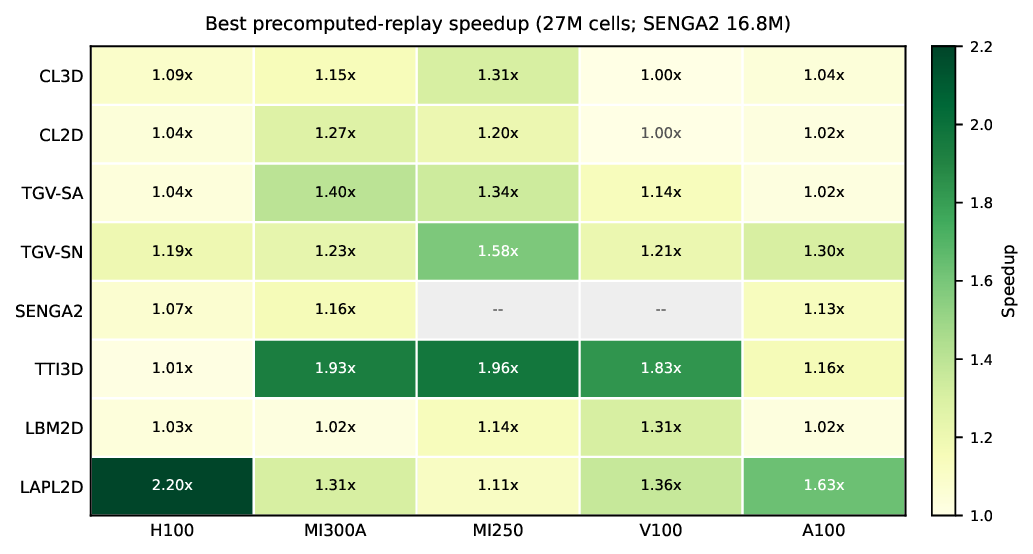}
      \caption{Best observed precomputed-replay speedup over the default OPS launch
  configuration for each application/GPU pair at approximately 27 million cells,
  except SENGA2 at 16.8 million cells.}
      \label{fig:speedup-overview-app-gpu}
  \end{figure}

  As shown in Figure~\ref{fig:speedup-overview-app-gpu}, we preview
  application-level behavior by selecting, for each application and GPU, the
  input configuration with the highest observed precomputed-replay speedup among
  evaluated inputs with approximately 27 million cells, except SENGA2 at
  16.8 million cells. This is not a global
  summary; it shows where tuned per-kernel launch shapes can improve performance
  across applications and GPU architectures. The complete distribution is
  analyzed in the following sections. The preview highlights that the largest
  visible gains concentrate in LAPL2D and selected 3D cases, while several
  application/GPU pairs show modest improvements near 1.0--1.1$\times$.

\subsection{Detailed Wall-Time and Speedup Results}

Table~\ref{tab:walltime-summary} reports median wall times for selected
representative GPU/application/input configurations. The speedup in this table
is computed only for that same configuration as $S=T_{\mathrm{Def}}/
T_{\mathrm{Pre}}$. Fixed-size aggregate speedup ranges for the same subset of
applications are reported separately in Table~\ref{tab:aggregate-speedups}.

\begin{table}[H]
  \centering
  \tiny
  \caption{Representative wall-time summary for specific input configurations.
  Def. is default OPS execution, AT is dynamic autotuning, and Pre. is
  precomputed replay. The Speedup $S$ is computed only for the shown configuration as
  Def./Pre.}
  \label{tab:walltime-summary}
  \setlength{\tabcolsep}{0.8pt}
  \renewcommand{\arraystretch}{0.78}
  \resizebox{0.84\columnwidth}{!}{%
  \begin{tabular}{@{}lllrrrr@{}}
  \toprule
  \textbf{GPU} & \textbf{App} & \textbf{Grid (cells)} &
  \textbf{Def.} & \textbf{AT} & \textbf{Pre.} & \textbf{S} \\
  \midrule
  H100 & CL3D   & $400{\times}300{\times}225$ (27.0M) & 7.60 & 12.52 & 7.15 & 1.063 \\
  H100 & TGV-SN & $450{\times}240{\times}250$ (27.0M) & 17.69 & 18.85 & 16.09 & 1.099 \\
  H100 & SENGA2 & $512{\times}512{\times}64$ (16.8M) & 144.31 & 150.54 & 139.93 & 1.031 \\
  H100 & TTI3D  & $300{\times}300{\times}300$ (27.0M) & 1.31 & 1.93 & 1.31 & 1.003 \\
  H100 & LBM2D  & $13500{\times}2000{\times}1$ (27.0M) & 12.71 & 12.68 & 12.32 & 1.032 \\
  H100 & LAPL2D & $9000{\times}3000{\times}1$ (27.0M) & 0.46 & 0.27 & 0.21 & 2.178 \\
  \midrule
  MI300A & CL3D   & $450{\times}240{\times}250$ (27.0M) & 16.01 & 34.61 & 14.39 & 1.112 \\
  MI300A & TGV-SN & $450{\times}200{\times}300$ (27.0M) & 47.40 & 54.05 & 42.87 & 1.106 \\
  MI300A & SENGA2 & $1024{\times}256{\times}64$ (16.8M) & 184.33 & 499.81 & 159.28 & 1.157 \\
  MI300A & TTI3D  & $450{\times}240{\times}250$ (27.0M) & 3.91 & 5.41 & 3.62 & 1.081 \\
  MI300A & LBM2D  & $3000{\times}9000{\times}1$ (27.0M) & 18.87 & 19.04 & 18.75 & 1.006 \\
  MI300A & LAPL2D & $5196{\times}5196{\times}1$ (27.0M) & 0.55 & 0.49 & 0.44 & 1.247 \\
  \midrule
  MI250 & CL3D   & $400{\times}300{\times}225$ (27.0M) & 25.59 & 56.18 & 24.74 & 1.034 \\
  MI250 & TGV-SN & $450{\times}250{\times}240$ (27.0M) & 90.62 & 88.56 & 67.33 & 1.346 \\
  MI250 & TTI3D  & $450{\times}240{\times}250$ (27.0M) & 7.13 & 8.47 & 5.11 & 1.395 \\
  MI250 & LBM2D  & $3000{\times}9000{\times}1$ (27.0M) & 23.41 & 22.84 & 21.32 & 1.098 \\
  MI250 & LAPL2D & $13500{\times}2000{\times}1$ (27.0M) & 0.97 & 0.94 & 0.87 & 1.113 \\
  \midrule
  V100 & CL3D   & $450{\times}240{\times}250$ (27.0M) & 43.46 & 64.05 & 42.96 & 1.012 \\
  V100 & TGV-SN & $360{\times}300{\times}250$ (27.0M) & 102.63 & 98.77 & 87.00 & 1.180 \\
  V100 & TTI3D  & $375{\times}300{\times}240$ (27.0M) & 9.77 & 7.55 & 5.37 & 1.818 \\
  V100 & LBM2D  & $3000{\times}9000{\times}1$ (27.0M) & 64.49 & 57.32 & 50.84 & 1.268 \\
  V100 & LAPL2D & $9000{\times}3000{\times}1$ (27.0M) & 1.13 & 0.91 & 0.84 & 1.347 \\
  \midrule
  A100 & CL3D   & $400{\times}300{\times}225$ (27.0M) & 16.81 & 22.83 & 16.35 & 1.028 \\
  A100 & TGV-SN & $450{\times}240{\times}250$ (27.0M) & 45.00 & 39.63 & 35.60 & 1.264 \\
  A100 & SENGA2 & $512{\times}512{\times}64$ (16.8M) & 273.30 & 272.32 & 262.71 & 1.040 \\
  A100 & TTI3D  & $360{\times}300{\times}250$ (27.0M) & 2.44 & 3.63 & 2.40 & 1.017 \\
  A100 & LBM2D  & $3000{\times}9000{\times}1$ (27.0M) & 21.09 & 21.13 & 20.75 & 1.017 \\
  A100 & LAPL2D & $2598{\times}10392{\times}1$ (27.0M) & 0.60 & 0.41 & 0.37 & 1.628 \\
  \bottomrule
  \end{tabular}
  }
  \renewcommand{\arraystretch}{1.0}
  \end{table}

\begin{table}[H]
  \centering
  \tiny
  \caption{Precomputed-replay speedup ranges at a fixed problem size for the
  representative application subset shown here. For each listed GPU/application
  pair, Min S, Median S, and Max S summarize $T_{\mathrm{Def}}/T_{\mathrm{Pre}}$
  across the input-shape configurations with the listed number of cells.}
  \label{tab:aggregate-speedups}
  \setlength{\tabcolsep}{1.8pt}
  \begin{tabular}{@{}lllrrr@{}}
  \toprule
  \textbf{GPU} & \textbf{App} & \textbf{Cells} &
  \textbf{Min S} & \textbf{Median S} & \textbf{Max S} \\
  \midrule
  H100 & CL3D & 27.0M & 1.051 & 1.069 & 1.090 \\
  H100 & TGV-SN & 27.0M & 1.040 & 1.113 & 1.195 \\
  H100 & SENGA2 & 16.8M & 1.028 & 1.032 & 1.073 \\
  H100 & TTI3D & 27.0M & 0.999 & 1.000 & 1.010 \\
  H100 & LBM2D & 27.0M & 1.029 & 1.030 & 1.032 \\
  H100 & LAPL2D & 27.0M & 2.168 & 2.168 & 2.180 \\
  \midrule
  MI300A & CL3D & 27.0M & 1.023 & 1.073 & 1.146 \\
  MI300A & TGV-SN & 27.0M & 1.048 & 1.110 & 1.234 \\
  MI300A & SENGA2 & 16.8M & 1.099 & 1.146 & 1.159 \\
  MI300A & TTI3D & 27.0M & 0.874 & 1.106 & 1.928 \\
  MI300A & LBM2D & 27.0M & 1.006 & 1.007 & 1.009 \\
  MI300A & LAPL2D & 27.0M & 1.256 & 1.262 & 1.307 \\
  \midrule
  MI250 & CL3D & 27.0M & 1.027 & 1.079 & 1.307 \\
  MI250 & TGV-SN & 27.0M & 1.285 & 1.333 & 1.583 \\
  MI250 & TTI3D & 27.0M & 1.395 & 1.439 & 1.962 \\
  MI250 & LBM2D & 27.0M & 1.098 & 1.122 & 1.139 \\
  MI250 & LAPL2D & 27.0M & 1.104 & 1.109 & 1.113 \\
  \midrule
  V100 & CL3D & 27.0M & 0.909 & 0.942 & 1.004 \\
  V100 & TGV-SN & 27.0M & 1.157 & 1.176 & 1.209 \\
  V100 & TTI3D & 27.0M & 1.705 & 1.793 & 1.834 \\
  V100 & LBM2D & 27.0M & 1.269 & 1.278 & 1.299 \\
  V100 & LAPL2D & 27.0M & 1.339 & 1.346 & 1.362 \\
  \midrule
  A100 & CL3D & 27.0M & 1.024 & 1.035 & 1.042 \\
  A100 & TGV-SN & 27.0M & 1.211 & 1.268 & 1.301 \\
  A100 & SENGA2 & 16.8M & 1.027 & 1.039 & 1.134 \\
  A100 & TTI3D & 27.0M & 0.966 & 1.014 & 1.155 \\
  A100 & LBM2D & 27.0M & 1.016 & 1.017 & 1.019 \\
  A100 & LAPL2D & 27.0M & 1.621 & 1.631 & 1.632 \\
  \bottomrule
  \end{tabular}
\end{table}

Dynamic autotuning is often slower than default OPS and precomputed replay
because it includes online exploration, amplified by the reduced iteration
counts used for benchmarking. Table~\ref{tab:walltime-summary} shows that
precomputed replay skips this search in representative runs, reaching
2.178$\times$ for LAPL2D on H100 and 1.818$\times$ for TTI3D on V100.
Table~\ref{tab:aggregate-speedups} shows the corresponding ranges across input
shapes, with maxima of 2.180$\times$ for LAPL2D, 1.962$\times$ for TTI3D, and
1.583$\times$ for TGV-SN.

\subsection{Global Speedup Results}

For each application-grid-GPU case, we use the median speedup over five repeats.
Table~\ref{tab:speedup-distribution} groups the completed analyzed cases by GPU
and dimensionality; e.g., H100 3D aggregates the 3D application/input
configurations that pass the completion and memory filters on H100. V100 3D has
$N=100$ because SENGA2 is omitted on V100, CL3D contributes 20 analyzed
configurations, TGV-SA contributes 20, and TGV-SN and TTI3D contribute 30 each.

  \begin{table}[H]
  \centering
  \scriptsize
  \caption{Distribution of median precomputed-replay speedups. Entries are counts of application-grid-GPU cases.}
  \label{tab:speedup-distribution}
  \setlength{\tabcolsep}{2.5pt}
  \begin{tabular}{@{}llrrrrr@{}}
  \toprule
  \textbf{GPU} & \textbf{Dim.} & \textbf{N} &
  \textbf{Slow} & \textbf{0--20\%} &
  \textbf{20--50\%} & \textbf{$\geq$50\%} \\
  \midrule
  H100   & 2D & 80  & 0  & 50  & 0  & 30 \\
  H100   & 3D & 125 & 4  & 113 & 8  & 0  \\
  MI300A & 2D & 80  & 10 & 34  & 36 & 0  \\
  MI300A & 3D & 125 & 5  & 64  & 35 & 21 \\
  MI250  & 2D & 80  & 0  & 79  & 1  & 0  \\
  MI250  & 3D & 120 & 1  & 43  & 60 & 16 \\
  V100   & 2D & 80  & 21 & 9   & 50 & 0  \\
  V100   & 3D & 100 & 20 & 47  & 13 & 20 \\
  A100   & 2D & 80  & 0  & 50  & 0  & 30 \\
  A100   & 3D & 125 & 11 & 88  & 26 & 0  \\
  \bottomrule
  \end{tabular}
  \end{table}

  Tuned per-kernel launch shapes improve 923 of the 995 cases in
  Table~\ref{tab:speedup-distribution}; global median speedups are
  1.024$\times$ on A100, 1.036$\times$ on H100, 1.124$\times$ on MI250,
  1.178$\times$ on V100, and 1.183$\times$ on MI300A. H100 and A100 show no 2D
  regressions, MI300A and MI250 show stronger 3D gains, and V100 has the most
  regressions while still retaining many 20--50\% and above-50\% speedups.

  \begin{figure*}[!t]
    \centering
    \includegraphics[width=\textwidth]{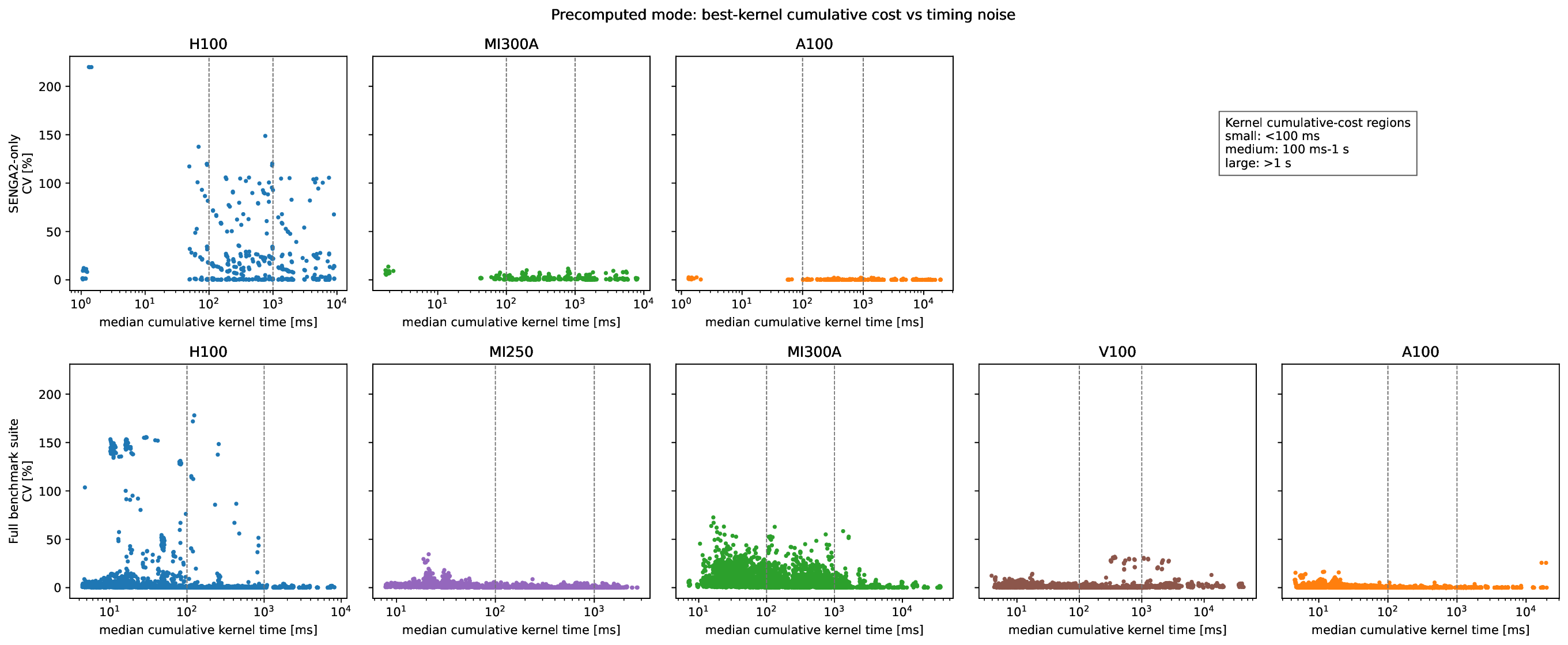}
    \caption{Kernel-level timing variability in precomputed replay mode. Each point
  represents one kernel, positioned by its cumulative execution time across all
  launches and its coefficient of variation (CV) across repeated runs.}
    \label{fig:kernel-runtime-cv}
  \end{figure*}

  \begin{figure*}[!t]
    \centering
    \includegraphics[width=1\textwidth]{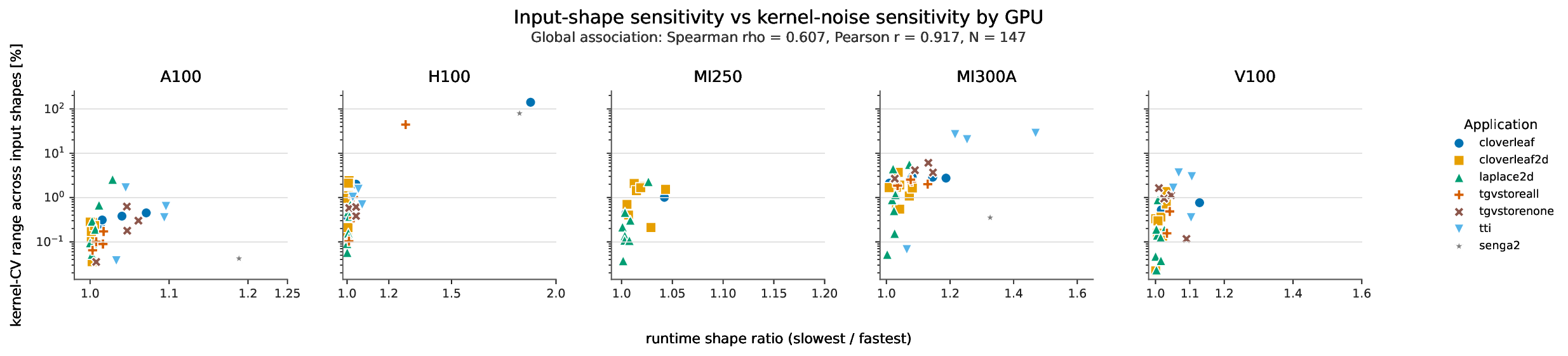}
    \caption{Input-shape sensitivity and kernel-level timing variability in precomputed replay mode. Each point represents one application, GPU, and fixed total cell
  count, evaluated across multiple grid-shape configurations. The x-axis shows the runtime shape ratio, computed as the slowest divided by the fastest precomputed
  runtime across those configurations, while the y-axis shows the range of kernel-level CV observed across the same input shapes.}
    \label{fig:kernel-shape-sens}
  \end{figure*}

\subsection{Impact of Input Problem Shape on Runtime}

  Input problem shape can affect runtime even at a fixed cell count.
  Table~\ref{tab:grid-order-effect} compares precomputed-replay runs with the
  same total cell count but different input shapes. The reported ratio is the
  slower runtime divided by the faster runtime, not a speedup over default,
  non-autotuned OPS.

  \begin{table}[H]
  \centering
  \scriptsize
  \caption{Representative input problem shape effects in precomputed replay. Fast and slow grids have the same total cell count with different input shapes. Times are medians over five repeats.}
  \label{tab:grid-order-effect}
  \setlength{\tabcolsep}{1.5pt}
  \resizebox{\columnwidth}{!}{%
  \begin{tabular}{@{}llccrrr@{}}
  \toprule
  \textbf{GPU} & \textbf{App} &
  \textbf{Fast grid} & \textbf{Slow grid} &
  \textbf{Fast (s)} & \textbf{Slow (s)} & \textbf{Ratio} \\
  \midrule
  H100   & CL3D  & $400{\times}160{\times}125$  & $250{\times}160{\times}200$  & 2.81  & 3.33  & 1.184 \\
  H100   & TGV-SN (OpenSBLI) & $400{\times}300{\times}225$ & $450{\times}300{\times}200$ & 15.49 & 16.16 & 1.043 \\
  H100   & TGV-SA (OpenSBLI) & $375{\times}300{\times}240$ & $450{\times}300{\times}200$ & 21.37 & 22.07 & 1.033 \\
  H100   & SENGA2 & $256{\times}256{\times}256$  & $1024{\times}512{\times}32$ & 128.35 & 166.97 & 1.301 \\
  MI300A & CL3D  & $640{\times}200{\times}500$  & $640{\times}400{\times}250$  & 28.79 & 32.55 & 1.131 \\
  MI300A & TGV-SN (OpenSBLI) & $375{\times}240{\times}300$ & $450{\times}400{\times}150$ & 39.72 & 44.83 & 1.129 \\
  MI300A & TGV-SA (OpenSBLI) & $375{\times}300{\times}240$ & $450{\times}200{\times}300$ & 41.52 & 44.55 & 1.073 \\
  MI300A & SENGA2 & $256{\times}256{\times}256$ & $1024{\times}512{\times}32$ & 141.18 & 190.08 & 1.346 \\
  MI250  & CL3D  & $375{\times}240{\times}300$  & $360{\times}300{\times}250$  & 22.77 & 29.24 & 1.284 \\
  V100   & LBM2D & $2263{\times}3536{\times}1$  & $3536{\times}2263{\times}1$  & 12.47 & 13.57 & 1.088 \\
  A100   & TTI3D & $640{\times}200{\times}500$  & $640{\times}500{\times}200$  & 5.52  & 6.04  & 1.094 \\
  A100   & TGV-SN (OpenSBLI) & $400{\times}300{\times}225$ & $375{\times}300{\times}240$ & 34.22 & 35.83 & 1.047 \\
  A100   & TGV-SA (OpenSBLI) & $450{\times}400{\times}150$ & $300{\times}300{\times}300$ & 48.66 & 49.04 & 1.008 \\
  A100   & SENGA2 & $256{\times}256{\times}256$ & $1024{\times}512{\times}32$ & 246.46 & 293.10 & 1.189 \\
  \bottomrule
  \end{tabular}
  }
  \end{table}

  Changing only the input problem shape while preserving the same total cell
  count can change precomputed-replay runtime by 0.8--34.6\% in these
  representative cases. The largest listed effect occurs for SENGA2 on MI300A,
  while SENGA2 also shows 30.1\% on H100 and 18.9\% on A100. These are
  shape-induced slowdowns, reinforcing that launch-shape tuning and input
  geometry interact. This table isolates input shape at fixed cell count; the
  present summary therefore does not identify problem size alone as the dominant
  factor across applications.

\subsection{Coefficient of Variation Analysis}

  We quantify timing noise using the coefficient of variation,
  $CV=\sigma/\mu$, computed from repeated precomputed-replay measurements.
  Figure~\ref{fig:kernel-runtime-cv} relates this variability to cumulative
  kernel runtime on a logarithmic scale. Short-running kernels can exhibit large
  relative CV because fixed launch, synchronization, scheduling, and
  timer-resolution overheads account for a larger fraction of their measured
  time; however, such kernels have limited impact unless they also contribute
  substantially to total runtime.

  At the application level, input problem shape and GPU/backend behavior also
  modulate runtime. Table~\ref{tab:grid-order-effect} shows that changing only
  the input problem shape can produce different runtime ratios across GPUs; for
  example, SENGA2 reaches 1.301 on H100, 1.346 on MI300A, and 1.189 on A100.
  Figure~\ref{fig:kernel-shape-sens} further compares runtime shape ratio with
  kernel-CV range, defined as the spread of kernel-level CV observed across the
  input-shape variants in each application/GPU/problem-size group. The positive
  association is moderate under Spearman correlation ($\rho=0.607$, $N=147$),
  with a strong Pearson correlation ($r=0.917$).

  This does not imply that longer kernels directly increase noise. Longer-running
  kernels can amortize fixed launch and measurement overheads, reducing relative
  CV. Instead, the results suggest that timing variability is driven by the
  interaction between kernel runtime contribution, input problem shape, and
  GPU/backend behavior.

\section{Conclusions and Future Work}

We have shown that GPU block/workgroup shape has a significant impact on
structured-mesh application performance, and that the best choice depends on the
kernel, input problem shape, and GPU architecture. The replay experiments isolate
the benefit of tuned block/workgroup shapes from online-search overhead,
demonstrating that launch-configuration tuning can improve OPS application
performance below the user-code level.

The results also show that performance sensitivity is not limited to the selected
launch shape. At fixed cell count, changing the input problem shape can produce
measurable runtime dispersion, and this sensitivity varies across applications
and GPUs. Figure~\ref{fig:kernel-shape-sens} further shows that input-shape
sensitivity and kernel-level CV are related descriptively, but not causally:
larger shape ratios tend to coincide with a wider spread of kernel-level timing
noise. Together with Figure~\ref{fig:kernel-runtime-cv}, this indicates that
timing stability depends on both runtime contribution and how the input problem
shape maps onto the GPU/backend. Thus, speedup, input-shape sensitivity, and
timing stability should be considered together when interpreting small
performance differences.

Future work will extend autotuning by re-evaluating selected block/workgroup shapes multiple times and estimating confidence intervals for their measured runtimes. This will enable confidence-
  aware rankings that distinguish consistently good configurations from those that appear optimal only because of measurement noise. Combined with runtime-extracted kernel metadata, iteration-space
  geometry, stencil-access information, target-GPU characteristics, input-shape descriptors, global CV summaries, and the Spearman association between input-shape sensitivity and kernel-CV range,
  these records can support models that predict near-optimal block/workgroup shapes without exhaustive online tuning. As demonstrated by Cui and Feng~\cite{cui2021iterml}, machine learning can
  iteratively prune and tune a dimensional parameter space, providing a basis for Intelligent Tuning. However, our work will focus on learning domain-decomposition policies that account for
  which kernels dominate runtime, how problem shapes map to each GPU/backend, and how work should be distributed across GPUs in large-scale HPC runs. Unlike Cui and Feng~\cite{cui2021iterml}, our
  future work will consider workloads distributed across multiple GPUs.

\section*{Acknowledgements}
This research was supported by the National Research, Development and
  Innovation Fund of Hungary (FK 145931), under the FK 23 funding scheme.
  We acknowledge Digital Government Development and Project Management Ltd.
  for awarding us access to the Komondor HPC facility in Hungary. We also
  acknowledge the EuroHPC Joint Undertaking and the LUMI consortium for
  providing access to the LUMI supercomputer.
  We used Gemini as AI tool for the Image creation of Figure~\ref{fig:ops-workflow-tuning},  that was already revised

\end{document}